\documentstyle{mn} 
\input psfig.sty
\title[$\Delta Y/\Delta Z$ from fine structure in the Main Sequence]
{$\Delta Y/\Delta Z$ 
from fine structure in the Main Sequence based on Hipparcos parallaxes 
\thanks{Based on data from the ESA Hipparcos astrometry satellite}}

\author[B.E.J. Pagel \& L. Portinari]
{B.E.J. Pagel$^{1}$ \& L.Portinari$^{1,}\,^{2}$\\  
\\$^{1}$NORDITA, Blegdamsvej 17, DK-2100 Copenhagen \O , Denmark\\  
$^{2}$Dipartimento~di~Astronomia,
Vicolo dell'Osservatorio 5, 35122 Padova, Italy}  
\date{1997 November}

\begin{document}
\maketitle 

\begin{abstract}
The slope $\Delta Y/\Delta Z$ is a quantity of interest in relation to
stellar evolution, the initial mass function and the determination of the
primordial helium abundance. In this paper we  
estimate  $\Delta Y/\Delta Z$ from fine structure  
in the Main Sequence of nearby stars  from
Hipparcos data for stars with $Z \leq Z_{\odot}$ and find a value of about 
3, which is consistent with what has been found in extragalactic H II
regions and with stellar models for suitable upper limits to the
initial masses of supernovae according to the IMF slope adopted.   
\end{abstract}

\begin{keywords}
galaxies: abundances, ISM: abundances, stars: abundances,
stars: evolution 
\end{keywords}

\section{Introduction} 

The ratio $\Delta Y/\Delta Z$ of fresh helium supplied to the interstellar
medium by stars relative to their supply of heavy elements is a 
quantity of considerable interest from several points of view.  It is a  
test of theoretical stellar yields (combined with the initial mass function) 
and it governs the slope of the regression of helium against oxygen in 
extragalactic H II regions and hence affects the use of that regression 
to determine the primordial helium abundance (Peimbert \& Torres-Peimbert
1976). From extragalactic H II regions, Lequeux et~al.\ (1979) found a
value of 3, combined with a primordial helium abundance $Y_{{\rm P}} = 
0.23$, and Pagel et~al.\ (1992) found a raw value of 6 with a similar 
$Y_{{\rm P}}$; Pagel et~al.\ applied various corrections to their raw
value of $\Delta Y/\Delta Z$ which reduced it to $4\pm 1$.  More recently,
Izotov, Thuan \& Lipovetsky (1997) have determined a higher value of 
$Y_{{\rm P}}=0.24$, which is in better accord with Big-Bang nucleosynthesis
theory with a low primordial value of D/H = $3.2\times 10^{-5}$ 
(Tytler 1997), and a lower value of the slope, $\Delta Y/\Delta Z\simeq 2$,
which is also closer to theoretical estimates if stars with initial
masses up to 50$M_{\odot}$ or more undergo supernova explosions as opposed to
collapsing into black holes (Maeder 1992).  Various determinations of 
$\Delta Y/\Delta Z$ from H II regions and planetary nebulae have been 
reviewed by Peimbert (1995), who concludes that the data are consistent
with the theoretical value of about 2.5, 
for the full range of the Scalo (1986) or Kroupa, 
Tout \& Gilmore (1993) IMFs.

An independent approach to the $\Delta Y/\Delta Z$ question is the
investigation 
of fine structure in the stellar zero-age Main Sequence, i.e.\ the 
dependence on $Z$ of the Main Sequence location.
Since concomitant changes
in $Y$ and $Z$ push the sequence in opposite directions,
for a given range of metallicities the related Main Sequences are more spread
apart if the corresponding variation in $Y$ is lower, or conversely
the broadening of the Main Sequence is a decreasing function of 
$\Delta Y/\Delta Z$
(Faulkner 1967; Perrin et~al.\ 1977; Pagel 1995; Cayrel de Strobel 
\& Crifo 1995; Fernandes, Lebreton \& Baglin 1996).  Perrin et~al., using 
ground-based parallaxes, could find no $Z$-dependence of the Main Sequence
location
in the luminosity--effective temperature plane for relatively metal-rich 
stars and deduced a large but very uncertain value: $\Delta Y/\Delta Z
\simeq 5\pm 3$.  Fernandes et~al., using Geneva colours and ground-based 
parallaxes, but without individual metallicity data, studied the scatter
in the $M_{V},\;B_{2}-V_{1}$ plane and just obtained a lower limit
$\Delta Y/\Delta Z > 2$ corresponding to the case in which all the scatter 
is assumed to be real. Cayrel de Strobel \& Crifo (1995) quote preliminary 
absolute helium abundances for three well studied visual binary stars 
with $Z$-values ranging from 0.2 to 1.7 times solar which are all 
consistent with a value of $\Delta Y/\Delta Z$ between 2 and 3.    
In this paper we use HIPPARCOS parallaxes (ESA 1997)
and infra-red flux temperatures (Alonso et~al.\ 1996) for a sample of mainly 
metal-poor stars with a view to improving on the previous 
results. Some preliminary results of this investigation, based on fewer 
stars, are reported by H\o g et~al.\ (1997).

\section{Theory} 

As has been discussed previously by Perrin  et~al.\ (1977) and 
Fernandes et~al.\ (1996), the effect of metallicity on the 
location of the stellar Main Sequence can be seen from quasi-homology
relations of the form (Cox \& Giuli 1968; Fernandes et~al.\ 1996)
\begin{equation}
\frac{L}{f(T_{{\rm eff}})} \propto \epsilon_{0}^{0.32}\kappa_{0}^{0.35}
\mu^{-1.33},
\end{equation}
where the energy generation constant $\epsilon_{0}\propto X^{2}$, the
opacity constant $\kappa_{0}\propto (1+X)(Z+Z_{0})$ with $Z_{0}\simeq
0.01$ (but see Section 5 below) and the molecular weight 
$\mu\propto (3+5X-Z)^{-1}$ leading to a
magnitude offset above the zero-age, zero-metallicity Main Sequence
where $X=X_{0}\simeq 0.76$
\begin{eqnarray}
 - \Delta M_{{\rm bol}}&\simeq &1.6\log\left[1-\frac{Z}{X_{0}}
 \left(1+\frac{\Delta Y} {\Delta Z}\right)\right] \nonumber \\
&&+0.87\log\left[1-\frac{Z}{1+X_{0}}\left(1+\frac{\Delta Y}{\Delta Z}
\right)\right] \nonumber\\
&&+0.87\log\left(1+\frac{Z}{Z_{0}}\right) \nonumber\\
&&+3.33\log\left[1-\frac{5Z}{3+5X_{0}}\left(1.2+\frac{\Delta Y}{\Delta Z}
\right)\right].
\end{eqnarray}
For high metallicities, around $0.7 Z_{\odot} \leq Z \leq 1.5 Z_{\odot}$,
the effects of $Y$ and $Z$ cancel out for $\Delta Y/\Delta Z\simeq 5.5$
(Fernandes et~al.\ 1996), but this is not the case for lower
metallicities (e.g. Faulkner 1967; Cayrel 1968).  
In the case of an old stellar population, direct application of Eq.~(2)
is also not very useful
in practice (as well as not being very accurate) because the effects of
stellar evolution increase sharply with luminosity above, say,
$M_V \simeq 5.5$, so that the sequences
cannot be expected to run straight and parallel over a wide range of
luminosities.  It is more useful to translate Eq.~(2) into a range of
$\log\,T_{{\rm eff}}$ at a fixed absolute magnitude using the slope of
the evolved Main Sequence, which is about 20 magnitudes per dex in
$T_{{\rm eff}}$.  We thus derive
\begin{eqnarray}
-\Delta\log\,T_{{\rm eff}}\!&\simeq &\!0.08\log\left[1-\frac{Z}{X_{0}}\left(1+
\frac{\Delta Y}{\Delta Z}\right)\right]\nonumber\\
&&\!+0.0435\log\left[1-\frac{Z}{1+X_{0}}\left(1+\frac{\Delta Y}{\Delta Z}
\right)\right] \nonumber\\
&&\!+0.0435\log\left(1+\frac{Z}{Z_{0}}\right) \nonumber\\
&&\!+0.167\log\left[1\!-\!\frac{5Z}{3+5X_{0}}\left(1.2\!+\!\frac{\Delta Y}{\Delta Z}
\right)\right].
\end{eqnarray} 
Since the terms in $Z$ are small, apart from $(1+Z/Z_{0})$, we can expand
the logarithms, also making use of $X_{0}\simeq 0.76$, to give
\begin{eqnarray} 
\Delta\log\,T_{{\rm eff}}&\simeq & 0.01Z+0.11Z(1+\Delta Y/\Delta Z) \nonumber\\ 
&&-0.044\log(1+Z/Z_{0}).
\end{eqnarray}
The quasi-homology fitting formula given by Faulkner (1967):
\begin{equation}
L\propto(X+0.4)^{2.67}(Z+Z_{0})^{0.455}\,f(T_{{\rm eff}})
\end{equation}
leads to a very similar relation
\begin{equation}
\Delta\log\,T_{{\rm eff}}\simeq 0.124Z(1+\Delta Y/\Delta
Z)-0.057\log(1+Z/Z_{0}).  
\end{equation} 
Besides chemical composition, the location of the Main Sequence of single
stars also depends on:
the treatment of convection and the size of core convective regions
(for $M \geq 1.1 M_{\odot}$);
rotation (for $M \geq 1.4 M_{\odot}$);
and evolution, inducing stars to deviate from the Zero Age Main Sequence
(for $M \geq 0.9 M_{\odot}$).
As discussed by Fernandes et~al.\ (1996), for magnitudes fainter
than $M_V \sim 5.5$ all the above effects are negligible,
and the broadening of the Main Sequence only depends on
chemical composition, namely $Z$ and  $\Delta Y/\Delta Z$.

Therefore, Eqs.~(4) and (6) provide a rough guide to the behaviour
of numerically
computed isochrones for magnitudes fainter than $M_{V}\simeq 5.5$ as a function
of $\Delta Y/\Delta Z$; qualitatively, the spread in the Main Sequence is
a decreasing function of this parameter. 

\section{The data} 
\begin{center} 
\begin{table*}
\begin{minipage}{150mm}
\caption{Data for the sample}
\label{tab1} 
\begin{tabular}{r r r r r r r r r r l}

\hline
\multicolumn{1}{c}{HIC} & \multicolumn{1}{c}{HD} & \multicolumn{1}{c}{$V$} &
\multicolumn{1}{c}{$M_{V}$} & \multicolumn{1}{c}{$\pm$} & 
\multicolumn{1}{c}{$T_{{\rm eff}}$} & \multicolumn{1}{c}{[Fe/H]} & 
\multicolumn{1}{c}{log $Z$} & \multicolumn{1}{l}{Notes} \\
 & \multicolumn{1}{c}{(or others)} \\
\hline
    5031 &   6348      &  9.15 & 6.18 & 0.10 & [4998] & [-0.67] & -2.31 \\
    5336 &   6582      &  5.17 & 5.78 & 0.03 &  5315  &  -0.67  & -2.31 &
    $\mu$ Cas\\
   10138 &  13445      &  6.12 & 5.93 & 0.01 & [5067] & [-0.24] & -1.99 \\
   11983 & \multicolumn{1}{r}{\scriptsize G 73-67} &  9.81 & 6.80 & 0.16 &  4756  &  -0.41  & -2.13 \\
   16404 & \multicolumn{1}{r}{\scriptsize G 246-38} &  9.91 & 6.14 & 0.19 &  5279  &  -2.94  & -4.38 \\
   17666 &  23439      &  7.67 & 5.72 & 0.12 & [4892] & [-1.03] & -2.43 & binary \\
   18915 &  25239      &  8.51 & 7.17 & 0.04 &  4842  &  -1.64  & -3.04 \\
   19849 &  26965      &  4.43 & 5.91 & 0.01 &  5040  &  -0.17  & -1.94 &
   o$^{2}$ Eri \\
   23080 &  31501      &  8.19 & 5.59 & 0.08 & [5254] & [-0.33] & -2.06 \\
   38541 &  64090      &  8.31 & 6.05 & 0.07 &  5441  &  -1.82  & -3.22 \\
   38625 &  64606      &  7.43 & 6.01 & 0.08 & [5070] & [-0.97] & -2.40 & binary \\
   39157 &  65583      &  6.99 & 5.87 & 0.03 &  5242  &  -0.60  & -2.27 \\
   41269 & \multicolumn{1}{r}{\scriptsize G 51-10} & 10.10 & 7.02 & 0.17 &  4579  &  -0.04  & -1.79 \\
   57939 & 103095      &  6.44 & 6.63 & 0.02 &  5029  &  -1.35  & -2.75 &
   Gmb 1830\\
   58949 & 104988      &  8.16 & 5.59 & 0.07 & [5247] & [-0.23] & -1.98 \\
   62607 & 111515      &  8.15 & 5.55 & 0.07 & [5354] & [-0.81] & -2.37 \\
   70681 & 126681      &  9.28 & 5.71 & 0.16 &  5541  &  -1.98  & -3.39 \\
   72998 & 131653      &  9.51 & 6.05 & 0.16 &  5311  &  -0.50  & -2.18 & excluded \\
   73005 & 132142      &  7.77 & 5.88 & 0.03 &  5098  &  -0.55  & -2.22 \\
   74234 & 134440      &  9.45 & 7.08 & 0.11 &  4746  &  -1.53  & -3.25 & [$Z$]=[Fe/H] \\
   74235 & 134439      &  9.07 & 6.73 & 0.09 &  4974  &  -1.47  & -3.19 & [$Z$]=[Fe/H] \\
   81170 & 149414      &  9.60 & 6.18 & 0.16 &  4966  &  -1.34  & -2.74 & excluded \\
   94931 & \multicolumn{1}{r}{\scriptsize BD  +41 3306} &  8.84 & 6.10 & 0.07 &  5004  &  -0.42  & -2.13 \\
   98020 & 188510      &  8.83 & 5.85 & 0.10 &  5564  &  -1.80  & -3.20 \\
   99461 & 191408      &  5.32 & 6.41 & 0.01 & [4893] & [-0.32] & -2.06 \\
  104214 & 201091      &  5.20 & 7.49 & 0.03 &  4323  &  -0.05  & -1.80 &
  61 Cyg A\\
  106122 & 204814      &  7.93 & 5.56 & 0.05 & [5232] & [-0.28] & -2.02 \\
  106947 & \multicolumn{1}{r}{\scriptsize G 126-19} &  9.51 & 5.74 & 0.18 &  5267  &  -0.20  & -1.96 \\
  109067 & \multicolumn{1}{r}{\scriptsize G 18-28} &  9.55 & 6.21 & 0.16 &  5411  &  -0.84  & -2.38 & excluded \\
  111783 & \multicolumn{1}{r}{\scriptsize G 67-8} &  9.50 & 5.55 & 0.20 &  5273  &  -0.09  & -1.86 \\
  116351 & \multicolumn{1}{r}{\scriptsize G 190-34} &  9.05 & 5.66 & 0.13 &  5318  &  -0.01  & -1.74 \\
\hline
\hline

\end{tabular}
\end{minipage} 
\end{table*}
\end{center} 
For the present investigation, we have combined the sample studied by H\o g
et~al.\ (1997), based on a proposal submitted to HIPPARCOS by  
one of us (BEJP) in 1982, with additional stars from the HIPPARCOS 
catalogue for which effective temperatures measured by the infra-red flux
method (Blackwell et~al.\ 1990) are available from the work of 
Alonso et~al.\ (1996).  We have chosen stars for which the HIPPARCOS
parallaxes 
are accurate to better than $\pm 9$~\%; among these, we have further
selected the stars with absolute magnitudes fainter than
$M_{V}=5.5$, which are expected to be reliable indicators of
$\Delta Y/\Delta Z$ (see Section~2). The resulting sub-sample of stars
is listed in Table~1. Columns 3 to 5 list the apparent visual magnitudes,
the absolute visual magnitudes and the corresponding errors, respectively.

Column~6 lists the effective temperatures.
The infra-red flux temperatures from Alonso et~al.\ are very accurate,
judging from the agreement between
different infra-red wavelength bands, and we assume them to have an 
accuracy of $\pm 50$~K, at least twice as good as the more heterogeneous 
temperatures available in other literature.  
This precision corresponds to an error of about $\pm 0.1^{{\rm m}}$ in 
the location of the Main Sequence, similar to the errors in $M_{V}$.
For the remaining stars, we have taken effective temperatures and
metallicities (shown in square brackets in Table 1) from 
either or both of the catalogues by Cayrel de Strobel et~al.\ (1992)
and Carney et~al.\ (1994).  

An important factor in comparing stellar data with theoretical isochrones
is the relationship between the metallicity [Fe/H] and the heavy-element
mass fraction $Z$  (columns~7 and 8, respectively). 
In most cases we have used the formula by Salaris, Chieffi \& Straniero (1993) 
\begin{equation}
Z=Z_{1}(0.638 f_{\alpha}+0.362),
\end{equation}
where $Z_{1}$ is the solar $Z$ ($Z_{\odot}=0.019$) scaled according to
[Fe/H] and $f_{\alpha}$ is the factor by which oxygen and $\alpha$-particle
elements are enhanced relative to iron, taking $f_{\alpha}$ from Pagel \&
Tautvai\v{s}ien\.e (1995).  Thus for [Fe/H] $< -1$, $Z = 2 Z_{1}$.
However, in the case of HD 134439 and 134440 we take $Z = Z_{1}$, following
King (1997). We assume [Fe/H] values to have an accuracy of $\pm 0.1$~dex.

Binary stars, marked with ``binary'' in Table~1, were naturally excluded
from further analysis.

\section{Comparison with theoretical isochrones}  



We have used grids of isochrones with different combinations of metallicity
($Z$ = 0.0004, 0.001, 0.004, 0.008, 0.019) and $\Delta Y/\Delta Z$ 
(= 0, 2, 2.5, 3, 3.5, 4, 5, 6), computed from Padova evolutionary
tracks normalized to the Sun: the helium content $Y$ for solar metallicity
isochrones and the mixing length parameter are calibrated 
so that the luminosity and effective temperature of the Sun
are reproduced for a model star with appropriate age and metallicity
($Z_{\odot} = 0.019$). A shift of the isochrones by $-0.009$ in 
$\log T_{{\rm eff}}$ turned out to be necessary in our plot
to fit solar-metallicity, faint Main Sequence stars. 
This shift will not affect our results,
since for our purpose it is basically the spread of the isochrones that counts,
rather than their absolute position. In fact, owing to many uncertainties
in the physics of stellar models (description of convective regions,
mixing length parameter, model atmospheres, opacities and so forth)
our analysis is not intended to give absolute values for $Y(Z)$, but only
to determine $\Delta Y/\Delta Z$ differentially on the basis of the relative
separation of isochrones of different metallicity.

On the other hand, even differential effects might be sensitive to the mixing 
length parameter $\alpha$,
which in principle might not be fixed but could 
change with mass, metallicity and/or
evolutionary phase. In our case, faint Main Sequence stars have very
similar masses and are in the same evolutionary phase,
so we are left with only a possible dependence of $\alpha$ on $Z$.
Chieffi, Straniero \& Salaris (1995), from inspection of HR diagrams of
globular clusters, have suggested that the mixing-length
parameter $\alpha$ might increase systematically with metallicity, pushing
the Main Sequence downwards. If so,
this could introduce systematic overestimations of $\Delta Y/\Delta Z$
from shifts in the sequence. However, 
this mixing-length effect was predicted to be small for the lower Main
Sequence, while it is expected to increase
 quite sharply with luminosity, so that,  
for sufficiently faint stars, the use of isochrones with a
constant mixing length para\-meter appears to be justified.  
Some checks on the possible role of magnitude-dependent effects other 
than $\Delta Y/\Delta Z$ are carried out in Section 5 below.

In Figures~1 to 3 we display sets of low Main Sequence isochrones for
representative values of $\Delta Y/\Delta Z$, together with the stellar data.
This first comparison suggests that 
$\Delta Y/\Delta Z =0$ is excluded, because the spread of the isochrones is 
too great, whereas $\Delta Y/\Delta Z = 6$ is too large (i.e.\ the isochrones
are too close); $\Delta Y/\Delta Z = 3$ gives a better fit.
This can also be seen qualitatively from the location of
HD~19445 (the `star' symbol at $M_V \simeq 5.1$), a well-known old,
very metal-poor star ([Fe/H] $= -2.15$, Alonso et~al.\ 1996). Although
it is brighter than the ``useful'' range of magnitudes and likely affected by
evolution, we can compare its location in the plot with the most
metal-poor 13~Gyr isochrone of our set. Both the $\Delta Y/\Delta Z = 0$ and
$\Delta Y/\Delta Z = 6$ case seem to be too extreme, while a good fit to
this star is obtained with $\Delta Y/\Delta Z = 3$.


These impressions are, however, largely based on the extremes in the abundance
range of the data,
and the intermediate-metallicity stars are too scattered to allow
any choice of $\Delta Y/\Delta Z$ from mere inspection.  In the next
section we shall try to improve on these qualitative impressions by applying
a maximum-likelihood calculation based on an extension of the idea of
quasi-homology. 


\begin{figure}
\psfig{file=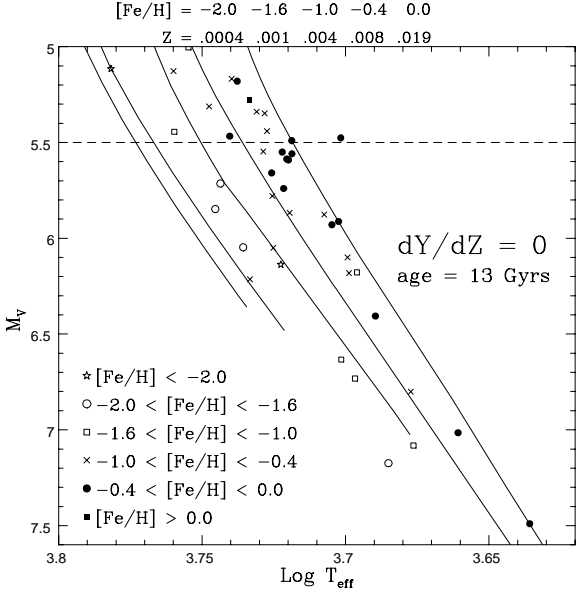,height=8.2truecm}
\caption{The stellar data from our sample are plotted against 13 Gyr Padova
isochrones calculated for $\Delta Y/\Delta Z =0$, shifted by -0.009 in 
$\log T_{{\rm eff}}$. The values of $Z$
for the 5 isochrones, together with the corresponding [Fe/H], are
indicated on top of the plot. The `star' symbol at $M_V \simeq 5.1$ indicates
HD~19445. The dashed line at $M_V = 5.5$ indicates the limiting magnitude for
the ``useful'' data listed in Table~1.
Binary stars are excluded from the plot.}
\end{figure}

\begin{figure}
\psfig{file=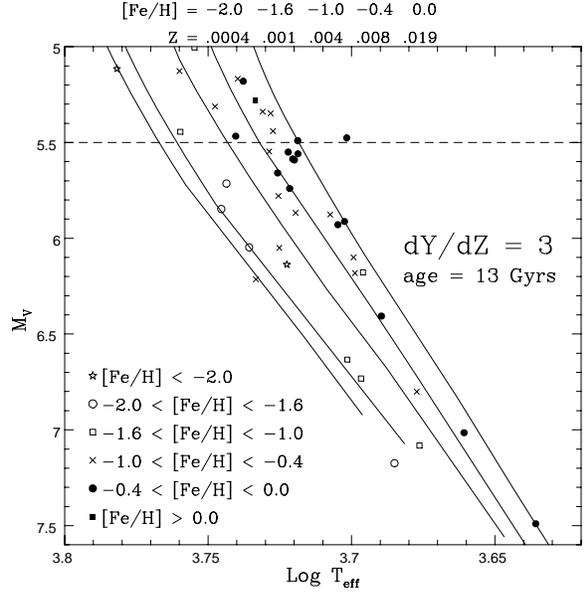,height=8.2truecm}
\caption{Same as Figure~1, but 
for $\Delta Y/\Delta Z = 3$.}
\end{figure}

\begin{figure}
\psfig{file=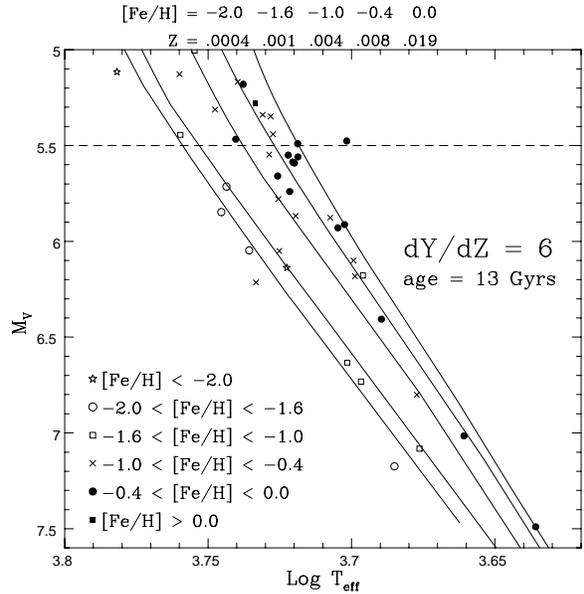,height=8.2truecm}
\caption{Same as Figure~1 and 2, but for $\Delta Y/\Delta Z = 6$.}
\end{figure}

\section{Statistical analysis using quasi-homology relations}

A quantitative analysis of the data is possible by
comparing them to theoretical relations like those discussed in Section~2.
Eqs.~(4) or (6) are not directly applicable, being
obtained on the simplifying assumption of a totally radiative structure,
while low Main Sequence stars actually have deep convective envelopes.

We have therefore established relations analogous to Eq.~(6) for our
numerically calculated isochrones of age 13~Gyr and in a limited range
of absolute magnitude, $5.5 \leq M_{V} \leq 7.5$.  Specifically, we
have imposed that our isochrones are described by a relation of the form
\begin{equation}
\phi(M_{V})\Delta\log\,T_{{\rm eff}}+k\log(1+Z/Z_{0}) =
a Z \left(1 + \frac{\Delta Y}{\Delta Z}\right),
\end{equation}
where $\phi(M_{V})$ is a normalization to allow for the convergence of the
isochrones towards low luminosities. 
Since in our set the solar-metallicity isochrone is fixed, while the
lower metallicity isochrones shift in $\log T_{{\rm eff}}$ according to $Z$ and
the assumed $\Delta Y/\Delta Z$, and moreover we do not know the reference
$\log T_{{\rm eff}} (Z=0)$, it is more 
convenient to re-formulate Eq.~(8) as:
\begin{eqnarray}
\phi(M_{V}) \Delta' \log\,T_{{\rm eff}} +
k \log \left(\frac{Z+Z_0}{Z_{\odot}+Z_0}\right) & = \nonumber \\
 = \, a (Z-Z_{\odot}) \left(1 + \frac{\Delta Y}{\Delta Z}\right), & \\
 & \nonumber \\
\Delta'\log\,T_{{\rm eff}}=
\log\,T_{{\rm eff}}(Z)-\log\,T_{{\rm eff}}(Z_{\odot}). & \nonumber
\end{eqnarray}
The scaling function $\phi(M_{V})$
and the suitable values of the parameters $k$, $a$ and $Z_0$ describing
the isochrones were calibrated
by means of numerical experiments using {\em Mathematica}. The specific
relation that we found is:

\begin{eqnarray}
 \frac{\Delta' \log\,T_{{\rm eff}}}{1-0.234(M_{V}-6.0)}
+ 0.054 \log \left(\frac{Z+Z_0}{Z_{\odot}+Z_0}\right) & = \nonumber\\ 
 = \, 0.147 (Z-Z_{\odot}) \left(1 + \frac{\Delta Y}{\Delta Z}\right); & \\
 & \nonumber \\
Z_{0} = 0.0015.~~~~~~~~~~~~~~~~~~~~~~~~~~~~~~~~~~~~~~~~~~~~ & \nonumber
\end{eqnarray}
The coefficients are quite similar to those in
Eq.~(6), but $Z_{0}$ turns out to be much lower than the widely quoted
value of 0.01.
\begin{figure}
\psfig{file=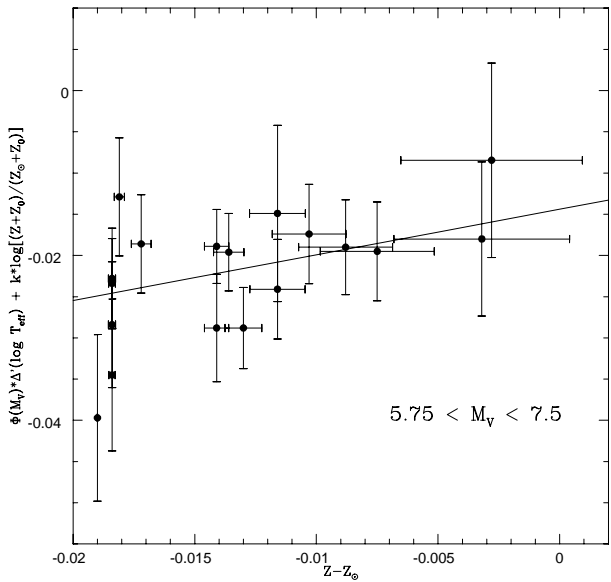,height=8.5truecm}
\caption{Maximum-likelihood regression for stars with $5.75\leq M_{V}\leq 7.5$,
assuming $\sigma_{T_{{\rm eff}}}=50$~K
and $\sigma_{{\rm [Fe/H]}}=0.1$. The slope corresponds to $\Delta Y/
\Delta Z =2.8 \pm 1.8$.} 
\end{figure}

We have calculated the expression on the left-hand side of Eq.~(10) for
the stellar data in our sample, and plotted it against $Z-Z_{\odot}$
with the corresponding errors on both axes. By means of a maximum-likelihood
fit computed
using the program of Pagel \& Kazlauskas (1992), we have derived the
experimental value of the slope $a (1+\Delta Y/\Delta Z)$ of
the regression line, and hence $\Delta Y/\Delta Z$.
Fig.~4  shows the plot for the data with $7.5 < M_V < 5.75$,
together with their maximum-likelihood fit.
In Table~2 we give maximum-likelihood solutions of this kind for
various ranges of absolute magnitude. We considered magnitude ranges from
$M_V = 7.5$ to brighter and brighter magnitudes, including one more star
at each step. This procedure allows us to estimate the limiting
magnitude where
evolutionary effects, or possibly the mixing length effects mentioned in
Section~4 which are expected to increase sharply with luminosity, begin
to play a role and spoil the determination of $\Delta Y/\Delta Z$. It was
further possible to isolate and exclude a few stars with large deviations
from our statistical analysis (marked with ``excluded'' in Table~1). It turns
out that all remaining stars fainter than $M_{V} \sim 5.75$ indicate a 
consistent
value of $\Delta Y/\Delta Z$. For stars brighter than this, 
the slope of the regression line rapidly increases displaying additional
effects, while for $M_{V}$ 
fainter than 6.0, the solutions become poor owing to shortage of stars. 
We therefore pay  most attention to the values determined
in the magnitude ranges between the two horizontal
lines in Table~2, and give a global estimate of
$\Delta Y/\Delta Z = 3\pm 2$ (s.e.) as the result of this investigation. 
\begin{center}
\begin{table} 
\caption{Max Likelihood solutions for $\Delta Y/\Delta Z$; 3 stars
excluded}
\label{tab2} 
\begin{tabular}{lcc} 
\hline
$M_{V}$ range & $\Delta Y/\Delta Z$ & N(stars)\\ \hline
7.5 to 7.00 & $2.3\pm 3.0$ &  4 \\
7.5 to 6.80 & $2.3\pm 3.0$ &  5 \\
7.5 to 6.70 & $3.7\pm 2.8$ &  6 \\
7.5 to 6.50 & $3.1\pm 2.7$ &  7 \\
7.5 to 6.40 & $3.0\pm 2.5$ &  8 \\
7.5 to 6.15 & $3.1\pm 2.5$ &  9 \\
7.5 to 6.12 & $4.4\pm 2.5$ & 10 \\
7.5 to 6.10 & $4.0\pm 2.4$ & 11 \\ \hline
7.5 to 6.00 & $3.6\pm 2.2$ & 12 \\
7.5 to 5.92 & $3.5\pm 2.1$ & 13 \\
7.5 to 5.90 & $3.2\pm 1.9$ & 14 \\
7.5 to 5.87 & $3.2\pm 1.9$ & 15 \\
7.5 to 5.85 & $3.2\pm 1.9$ & 16 \\
7.5 to 5.80 & $2.8\pm 1.8$ & 17 \\
7.5 to 5.75 & $2.8\pm 1.8$ & 18 \\ \hline   
7.5 to 5.72 & $3.5\pm 1.8$ & 19 \\  
7.5 to 5.70 & $4.0\pm 1.7$ & 20 \\
7.5 to 5.60 & $6.1\pm 1.5$ & 21 \\
7.5 to 5.59 & $6.1\pm 1.4$ & 22 \\
7.5 to 5.57 & $6.0\pm 1.4$ & 23 \\
7.5 to 5.55 & $6.0\pm 1.4$ & 25 \\
7.5 to 5.50 & $6.2\pm 1.4$ & 26 \\
\end{tabular}
\end{table} 
\end{center}


\vspace{-5mm} 
\section{Discussion}

Although we have used the best available data for as many stars as possible  
having absolute magnitudes and effective temperatures with 
sufficient accuracy, our result for $\Delta Y/\Delta Z$ still has a 
disappointingly large uncertainty.  Nevertheless we consider it an 
improvement on previous work, although a quite similar value ($\sim 3.5$) 
was actually 
estimated by Faulkner (1967) thirty years ago.  Our value is intermediate 
between those 
derived from extragalactic H II regions by Pagel et~al.\ (1992) and 
Izotov, Thuan \& Lipovetsky (1997) and also agrees well with the values 
quoted by Peimbert (1995), which supports the assumption that in this 
respect the stars in our Galaxy and in dwarf irregulars have undergone 
similar evolution, contrary to the suggestion that in the latter case 
$\Delta Y/\Delta Z$ has been increased by metal-enhanced galactic winds 
(e.g. Pilyugin 1993).  Furthermore, we see no evidence for significant 
variations in $\Delta Y/\Delta Z$ relative to the precision 
of currently available data.

Our result is also consistent with the conclusion of 
Fernandes et~al.\ (1996) that $\Delta Y/\Delta Z > 2$.  Taking that lower 
limit
together with the calculations of Maeder (1992), one finds upper mass limits 
for supernovae of $55 M_{\odot}$ for a Salpeter IMF and $100 M_{\odot}$
for a Scalo IMF.  A similar limit has been 
deduced on other grounds for a Salpeter-like IMF slope by Tsujimoto et~al.\ 
(1997).  On the other hand, the Miller-Scalo (1979) IMF would permit an
arbitrarily high upper mass limit for $\Delta Y/\Delta Z= 2.56$ (Traat
1995).  

\subsection*{Acknowledgments}

We thank Erik H\o g for supplying the HIPPARCOS parallaxes, L\'eo Girardi
for supplying Padova code evolutionary tracks, Peter Thejll for his 
assistance in dealing with the HIPPARCOS data, Jim MacDonald and John Faulkner
for stimulating discussions.  L.P. acknowledges financial support from the
Danish Rektorkollegiet and the Italian MURST, and thanks  
{\em NORDITA} for hospitality and additional
support during the preparation of this paper.

{

\end{document}